\documentclass[letterpaper]{mn2e}
\usepackage{epsfig,natbib2,natbibmnfix}

\newcommand{\be}{\begin{equation}}
\newcommand{\ee}{\end{equation}}

\newcommand{\apj}{ApJ}

\newcommand{\mnras}{MNRAS}
\newcommand{\aap}{A\&A}

\newcommand{\apjl}{ApJL}

\newcommand{\nat}{Nature}

\def\ltsima{$\; \buildrel < \over \sim \;$}
\def\simlt{\lower.5ex\hbox{\ltsima}}
\def\gtsima{$\; \buildrel > \over \sim \;$}
\def\simgt{\lower.5ex\hbox{\gtsima}}
\def\sgra{Sgr~A$^*$}

\def\msun{{\,{\rm M}_\odot}}
\newcommand\mbh{{\,{\rm M}_{\rm bh}}}

\def\del#1{{}}

\title{Forced accretion in stochastically fed AGN and quasars}

\author[S.~Nayakshin and A.~King] {\parbox{18cm}{Sergei
Nayakshin\footnotemark[1] and Andrew King}\vspace{0.3cm}\\ Department of
Physics \& Astronomy, University of Leicester, Leicester, LE1 7RH, UK}

\begin{document}

\maketitle

\begin{abstract}
Steady state accretion discs larger than $\sim 0.01-0.1$ pc are known to be
gravitationally unstable for the accretion rates needed to explain
super-massive black hole (SMBH) activity. We propose that SMBH are fed by a
succession of mass deposition events with randomly directed angular
momenta. Because of incomplete angular momentum cancellation a warped
accretion disc forms in the inner few parsec. The orientation of the disc
performs a random walk. Deposition of new material promotes SMBH accretion at
rates much faster than viscous. Observational implications of this picture
include: (i) lighter accretion discs that can fuel AGN and quasars and yet
avoid star formation at $R \gg 0.1$ pc; (ii) star formation inside the disc
is not a function of mass accretion rate only.  It can take place at
high or low accretion rates, e.g., when too few clouds arrive in the inner
region. An example of this might be the central parsec of our Galaxy. (iii)
The discs can form Compton-thick obscuring structures of $\sim$ parsec size as
required in AGN unification models; (iv) faster black hole growth resulting
from misalignment of the disc and the black hole spin in the early Universe;
(v) Isotropic deposition of SMBH energy and momentum feedback in the galaxy
bulge. This may help explain the high efficiency with which it seems to be
operating in the Universe. (vi) No correlation between SMBH activity and the
presence of kiloparsec scale bars or gaseous discs in galactic bulges; (vii)
Bodily collisions between gaseous components of merging galaxies facilitate
production of gas streams feeding the centre of the combined galaxy. Mergers
should thus be catalysts of SMBH growth. (viii) Conversely, galaxies
experiencing fewer mergers are more likely to form massive nuclear star
clusters than feed their SMBHs.
\end{abstract}

\begin{keywords}
{Galaxy: centre -- accretion: accretion discs -- galaxies: active}
\end{keywords}
\renewcommand{\thefootnote}{\fnsymbol{footnote}}
\footnotetext[1]{E-mail: {\tt Sergei.Nayakshin at astro.le.ac.uk}}

\section{Introduction}\label{sec:intro}

A well known difficulty in fuelling active galactic nuclei (AGN) is that the
typical angular momentum of gas in the galactic bulge is very large compared
with that of the last stable orbit around a black hole
\citep[e.g.,][]{Krolik99,Combes01,Jogee04}. Assuming that the bulge is the
reservoir ultimately supplying mass to the nucleus, the result is presumably a
disc with the size of a fraction of the bulge radius, i.e., a fraction of a
kiloparsec. The material at the inner edge of the disc would then have to give
up its angular momentum rapidly enough to be able to lower itself in the SMBH
potential well. This can perhaps be done through the action of stellar and
gaseous bars and other gravitational torques \citep[e.e.,][]{Shlosman90}.

However, there is no clear observational evidence for a link between AGN
activity and the presence of ``grand design'' gas discs or stellar bars
\citep{Combes03}.  A further theoretical impasse for large scale gaseous discs
is that even if material does reach the inner parsec scales, theory predicts
that these discs are too cold and too massive, and thus should be unstable to
gravitational fragmentation and star formation
\citep[e.g.,][]{Paczynski78,Kolykhalov80,Shlosman89,Collin99}. For SMBH
feeding, star formation is a disease that threatens the very existence of
accretion discs outside the ``self-gravity radius'', $R \simgt 0.01-0.1$ pc
\citep{Goodman03}. If gaseous discs are turned into stellar discs, SMBHs
cannot grow by gas accretion.

Faced with these and other theoretical difficulties, \cite{Goodman03} and
\cite{KingPringle07} suggested that AGN many may be fed by a direct deposition
of low angular material into the region $R < R_{\rm sg}$. The main difficulty
for this suggestion is that the specific angular momentum of the orbit at $R =
R_{\rm sg}$ is only $l \sim 100$ pc km/s, much smaller than the more typical
$l \sim 10^4-10^5$ pc km/s at $ R = 10^2 - 10^3$ pc.

On the other hand, the SMBH mass, though large, is only a tiny fraction($\sim
0.001$) of the stellar mass of the bulge \citep{Haering04}, and must be an
even smaller fraction of the original gaseous mass from which the bulge is
made. It thus seems not impossible that cloud-cloud collisions, supernova
shocks, and galaxy mergers channel such a small fraction of bulge gas on
nearly radial orbits ending up inside $R < R_{\rm sg}$. Unfortunately, a
quantitative treatment of these processes would be very model-dependent and
hardly constraining.

In this paper we note that gas feeding the SMBH does not actually need to have
such a small angular momentum after all. We propose that SMBHs can be fed by
clouds with angular momentum an order of magnitude higher, i.e.  $l \sim 10^3$
pc km/s or more, if there is significant cancellation of angular momentum by
shocks in the SMBH's vicinity. This would happen if the angular momentum of
the clouds is random. Below we show that such a picture predicts the formation
of a warped accretion disc that can be much larger than $R=R_{\rm sg}$, and we
discuss observational implications for current observations of AGN and for the
SMBH-galaxy connection. We feel that this ``stochastic cloud'' mode of SMBH
fuelling has numerous advantages over the ``grand design'' models.

\section{Can grand design accretion feed SMBH?}\label{sec:problems}

\cite{Goodman03} presented the most complete and up-to-date consideration of
gravitational stability of different accretion disc models. The standard
accretion flow model is gravitationally unstable beyond the self-gravity
radius $R_{\rm sg}$, which is a function of the accretion rate in the disc,
$\dot M$. A rough power-law fit to this dependence from Figure 1 in
\cite{Goodman03} is
\begin{equation}
R_{\rm sg} \approx 0.01 \;\hbox{pc}\; \dot M^{-2/7}\;,
\label{rsg}
\end{equation}
where $\dot M$ is in units of $\msun$~year$^{-1}$. For astrophysically
interesting accretion rates, then, $R_{\rm sg}\sim 0.01 - 0.1$~pc.
\cite{Goodman03} also found that none of the more sophisticated models avoids
becoming self-gravitating at parsec-scale distances from the SMBH, unless an
unspecified and very large energy source is applied to the disc to keep it hot
enough. The required energy at 10 pc, for example, was at least that expected
if all of the disc material was reprocessed by massive stars. \cite{Goodman03}
also pointed out that even if a source for this energy was found, the disc may
still be thermally unstable as the cooling times were much shorter than
the local dynamical time. Furthermore, \cite{Sirko03} demonstrated that such a
high energy liberation rate near an AGN is actually in conflict with the
observed spectra of typical AGN.

\cite{Thompson05} proposed a starburst disc model which appears to overcome
the difficulties noted by \cite{Goodman03}. These authors suggested
parameterising the radial inflow velocity at a fraction of the disc sound
speed, arguing that external torques or spiral density waves may deliver the
required angular momentum transfer. This makes these discs lighter at a given
accretion rate. Additionally, Thompson et al.'s star formation rate in the
disc is limited by the action of energy and momentum feedback from massive
stars and supernovae, allowing some fuel to trickle down all the way to the
nucleus.

Without numerical analysis it is difficult to say whether this model will work
for parsec-scale AGN discs. One problem is that of time scales.  Gravitational
collapse of the disc is expected to take place on the dynamical time scale,
\begin{equation}
t_{\rm dyn} = \frac{R^{3/2}}{G^{1/2}\mbh^{1/2}} \sim 10^{3} R_{\rm pc}^{3/2}
 M_8^{-1/2} \hbox{years}\;.
\label{tdyn}
\end{equation}
\cite{Thompson05} show that the disc cooling time is much shorter than the
dynamical time in their model. On the other hand, the lifetime of massive
stars is at least a few million years. Therefore, the supernovae contribution
to the feedback might be activated too slowly in this model, i.e., only when
the disc would already have collapsed gravitationally.

Feedback due to radiation and outflows from massive stars may appear
much faster, perhaps as quickly as required by equation \ref{tdyn}. However it
is not obvious that the feedback will be spread uniformly enough in the
disc. For a disk of scale height $H$, radiation and outflows from a massive
star would affect a surface area of size $\sim \pi H^2$ only. Thus, to heat up
the whole disc, $\sim (R/H)^2$ massive stars are required. If we assume that
one massive star forms from $\sim 500 \msun$ of gaseous material for a
\cite{Salpeter55} IMF, this would require disc mass of $M_{\rm disc} \sim 5\times
10^6 \msun h_{-2}^{-2}$, where $h_{-2}= 100 H/R$. The corresponding star
formation rate would be $M_{\rm disc}/t_{\rm dyn} \sim 5\times 10^3
\msun$~year$^{-1} h_{-2}^{-2} M_8^{1/2} R_{\rm pc}^{-3/2}$. This seems to be
too high for the central region of only a few parsecs.

In addition, analytical estimates \citep{NC05} and numerical simulations
\citep{NCS07} show that the stellar component of star forming discs decouples
vertically from the gaseous disc when the total stellar mass becomes
comparable to the gas mass. Since gas cools radiatively and stellar
motions do not, stellar discs become more vertically extended, as in the
Galactic disc in Solar neighbourhood.  Feedback in massive star-dominated discs
in AGN would thus occur mainly outside the gaseous disc, diminishing the
efficiency of feedback deposition further.

Summarising, there does not appear to be a completely convincing way to avoid
the self-gravity catastrophe for parsec-scale {\em massive} AGN discs. The
fundamental problem is that these discs are very inefficient as far as angular
momentum transfer is concerned. They need to be massive enough to provide a
high enough rate of SMBH feeding, yet they cannot be hot enough to escape
fragmentation. The simplest conclusion one can draw from this is that while
such discs are probably very important for transferring huge amounts of gas
into the inner parts of galaxies, they probably fail to fuel SMBHs.

\section{Stochastic SMBH feeding}\label{sec:stochastic}

\begin{figure*}
\begin{center}
\begin{tabular}{cc}
\leavevmode
\epsfxsize=0.55\textwidth\epsfbox{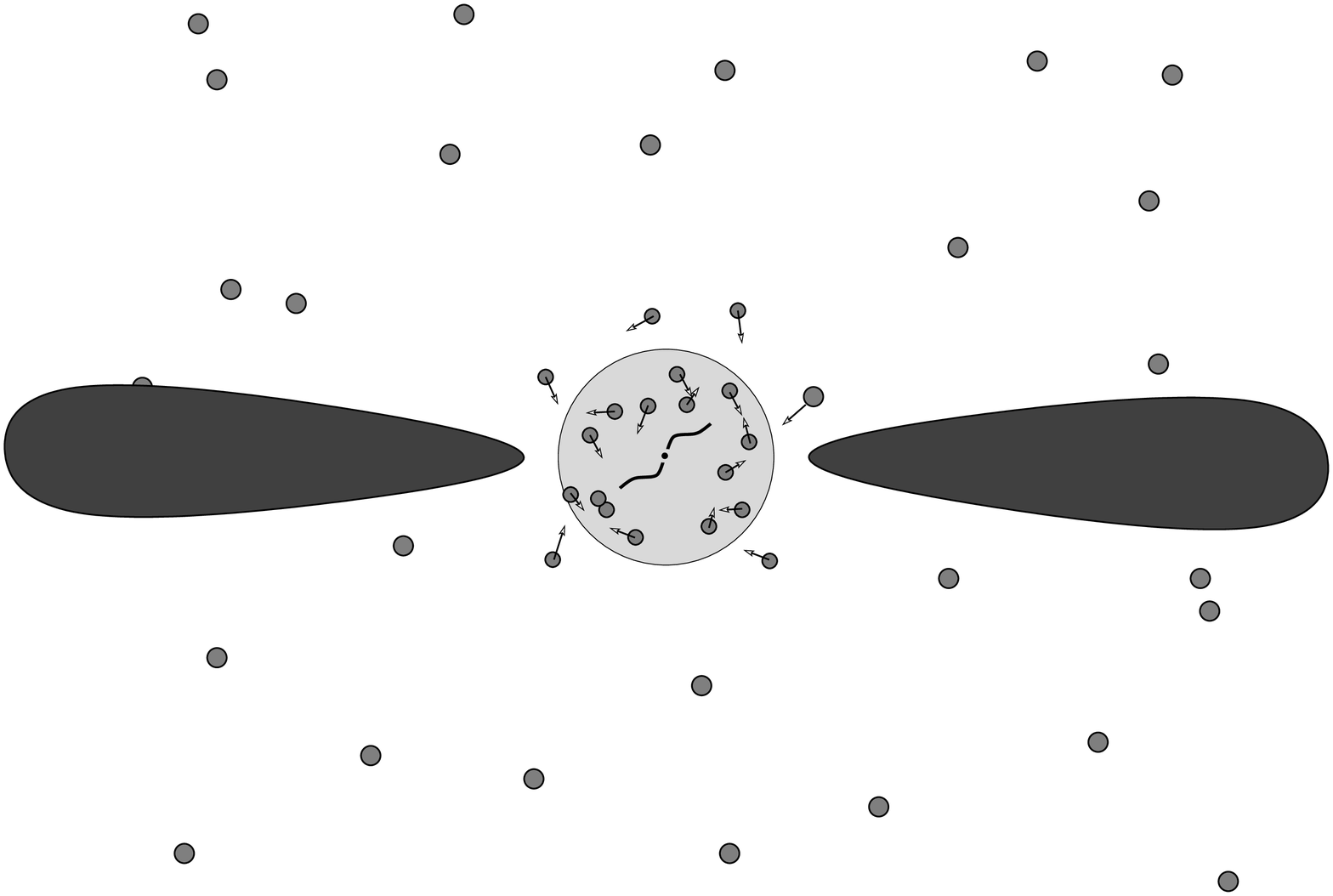} &
\epsfxsize=0.4\textwidth\epsfbox{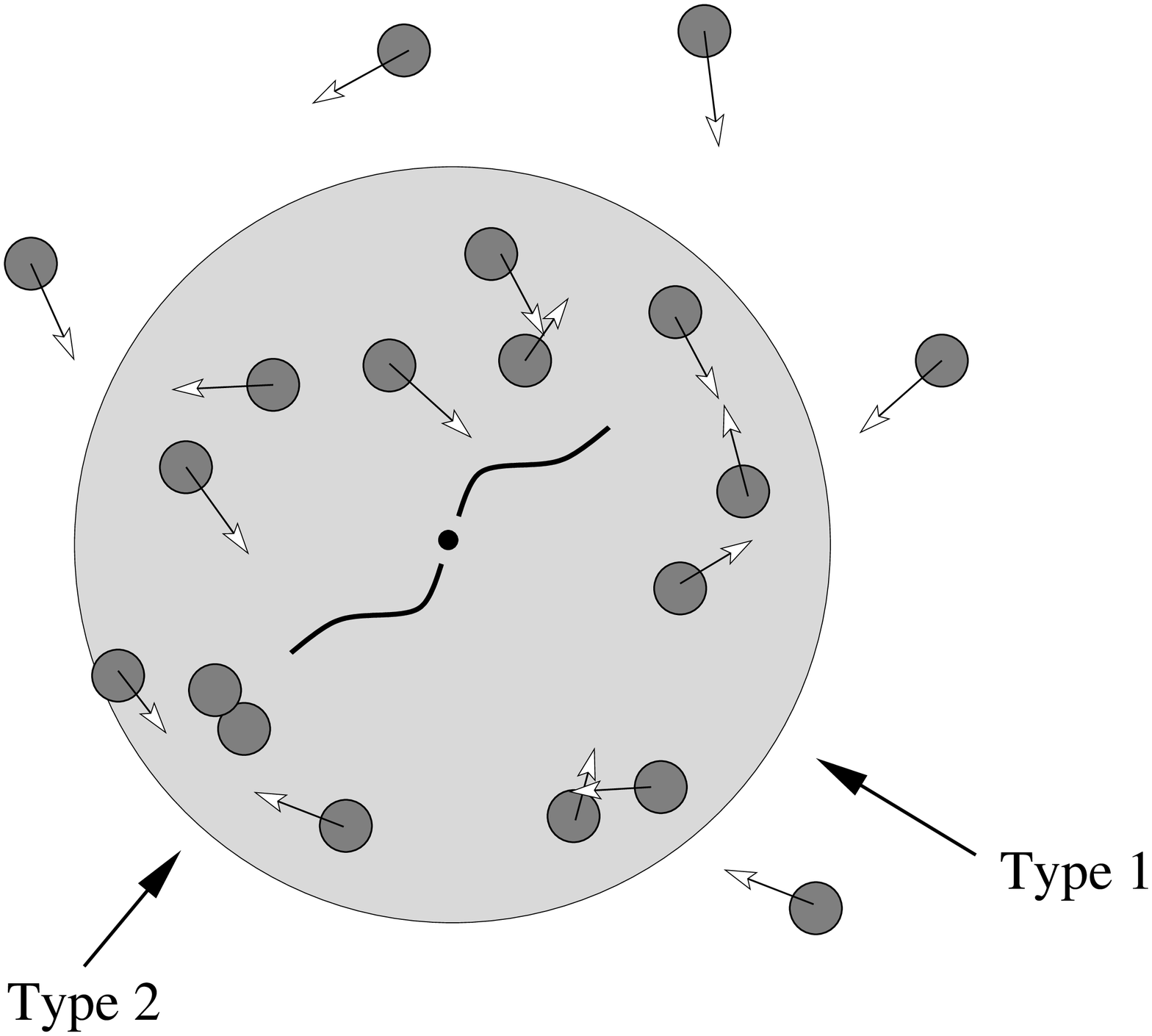} \cr
\end{tabular}
\end{center}
\caption{Schematic representation of  stochastic cloud AGN feeding. {\bf
  Left:}  large scale ($100$pc to 1 kpc) distribution of gas. Most of the gas
  has a significant angular momentum and is in a circularised large scale disc
  that contributes little to AGN feeding. The quasi-spherical halo of gas
  clouds produces clouds on small angular momentum orbits.  {\bf Right:}
  Zoom-in view of the central region (the inner parsecs). Clouds collide in
  the hatched ``collision sphere'', and form a warped disc whose orientation
  changes stochastically. Lines of view clear of the clouds and warped disc
  yield type 1 AGN classification whereas lines of sight intersecting the
  disc or clouds present AGN as a type 2 source. Note that this division
  into Type 1 and 2 is caused by the inner stochastically oriented disc and is
  unrelated to the large-scale structure of the galaxy.}
\label{fig:fig1}
\end{figure*}

\begin{figure}
\centerline{\psfig{file=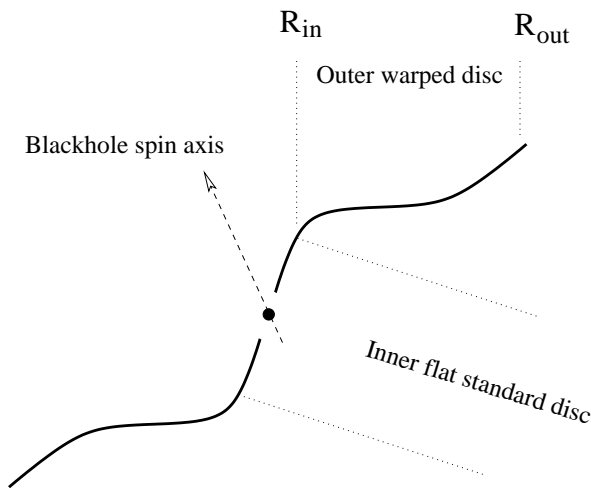,width=.45\textwidth,angle=0}}
\caption{Illustration of the accretion disc structure in Figure 1. The outer
  disc extends from $R_{\rm in} \sim 0.01$ pc to $R_{\rm out}$. Accretion is
  driven by cancellation of angular momentum in collisions between the disc
  and infalling clouds. The inner viscous disc joins the outer one at $R=
  R_{\rm in}$. A short viscous time keeps the inner disc flat and aligned with
  the inner fringes of the outer disc, but not necessarily aligned with the
  black hole spin. The outer disc is warped and can be highly disturbed,
  presenting a large solid angle for AGN illumination.}
\label{fig:fig2}
\end{figure}

\subsection{Large scale region}\label{sec:radial}

Figure 1 illustrates the schematics of the model we would like to explore
here. We assume that during the epoch of bulge and SMBH growth, the gaseous
medium feeding both of these can be crudely divided into two components (left
panel of figure 1). The first is the dominant part of the gas with 
non-negligible angular momentum. This component forms a large scale gaseous
disc that mainly forms stars for the reasons explained in \S
\ref{sec:problems}. The second component is a less massive quasi-spherical
one. A key assumption we make here is that these two components do not mix
completely, or else we would only have the gaseous disc. This immediately
implies that the quasi-spherical component must be in the form of compact
clouds\footnote{Alternatively, it can be a hot hydrostatically supported
gas. The gas cooling time must nevertheless be short to feed the SMBH at an
appreciable rate. This cooling gas would thus end up forming cool clouds
again.}. We reason further that there will be continuous production of
clouds on nearly radial trajectories through cloud-cloud collisions, their
gravitational scattering off each other or off star clusters, supernova driven
shock waves, etc. We now consider the structure of the inner accretion flow
assuming that these nearly radial clouds provide the dominant mechanism of
SMBH feeding.

\subsection{Collision sphere and disc formation}

The right panel of Figure 1 zooms in on the inner parsecs of the model. Let
$R_{\rm coll}$ be the radius of the sphere centred on the SMBH that is
``optically thick'' to a typical gas cloud. Clouds collide, convert their bulk
kinetic energy into heat which is radiated away rapidly, and become bound to
the SMBH inside this region.  The lower limit on $R_{\rm coll}$ is given by
the geometrical size of the infalling clouds themselves, as the different
sides of the cloud then collide with each other. Observed interstellar clouds
form structures on a range of scales, but scales of few parsec are most
common \citep{Elmegreen96}. We thus expect that $R_{\rm coll} > 1$~pc.

The angular momentum of the clouds at $R_{\rm coll}$ is at most the local
circular orbital value, $l_{\rm coll}=\Omega_{\rm coll} R_{\rm coll}$. As
clouds collide, they circularise and settle into a disc. The outer radius of
the disc, $R_{\rm out}$, is smaller than $R_{\rm coll}$, and is controlled by
the degree to which angular momentum is cancelled in these collisions.

To estimate it, define $\dot N$ as the average rate of number of clouds
entering the collision sphere $R= R_{\rm coll}$.  The angular momentum of the
material entering the sphere is then, by random walk arguments, $l_{\rm
feed}\sim l_{\rm coll}/N_{\rm feed}^{1/2}$, where $N_{\rm feed} = \dot N
\Omega_{\rm coll}^{-1}$.  Suppose that the collision sphere is within the
black hole sphere of influence, $R_{\rm bh} = G \mbh/\sigma^2$, where $\sigma$
is the bulge velocity dispersion. In that case the outer edge of the disc is
given by
\begin{equation}
R_{\rm out} = \frac{R_{\rm coll}}{N_{\rm feed}}\;.
\label{rout}
\end{equation}
Obviously, in the limit of $\dot N\rightarrow \infty$, we get $R_{\rm out}
\rightarrow 0$, as expected in the case of a zero angular momentum inflow.

\del{Now, if $R_{\rm out}$ is comparable or less than the self-gravity radius,
then the self-gravity catastrophe does not occur, and most of the fuel will
presumably make it into the SMBH in due course. However, as $R_{\rm sg}$ is
such a tiny region, and it is more likely that $R_{\rm out} \gg R_{\rm
sg}$. Therefore, the disc formed in the way described above might become
gravitationally unstable if it is massive enough.}

\subsection{Forced accretion in the outer accretion disc}\label{sec:forced}

The AGN disc we consider has a very unusual feeding mode as the angular
momentum of the material supplied to the disc fluctuates with arrival of new
clouds. Unlike binary Roche lobe overflow systems, the geometry is 3D, with
new material arriving not necessarily at the outer edge. In particular, parts
of clouds with particularly low angular momentum might strike the disc further
in than $R_{\rm out}$. In a clear difference from the standard accretion flow,
no angular momentum transport is needed in such randomly fed discs to promote
accretion. Cancellation of the randomly directed angular momentum of the new
material and that of the disc sets up a radial inflow. We call this ``forced
accretion'' as it is mediated by external deposition of material.

Let us estimate the steady-state mass of such a disc. The specific angular
momentum of a disc containing $N$ equal mass clouds that arrived with a random
direction of the angular momentum will be $l_{\rm disc} \sim l_{\rm
coll}/N^{1/2}$, again by random walk arguments. If this angular momentum is
less than $l_{\rm sg}$, the disc will move in radially (in this very simple
picture) to $R < R_{\rm sg}$. Thus, $N= (l_{\rm coll}/l_{\rm sg})^2$ random
cloud deposition events are needed to lower the gas to $R < R_{\rm sg}$
region. Accordingly, the ``stochastic average'' mass of the disc will be
\begin{equation}
M_{\rm disc} = M_1 N \sim M_1 \left(\frac{l_{\rm coll}}{l_{\rm sg}}\right)^2 =
\frac{M_1 R_{\rm coll}}{R_{\rm sg}}\;,
\label{mdisk}
\end{equation}
where $M_1$ is the mass of a cloud. For example, if $M_1 = 100 \msun$ and
$R_{\rm coll}/R_{\rm sg}=300$, then $M_{\rm disc}=3\times 10^4 \msun$. 
The accretion time scale is given by the expression
\begin{equation}
t_{\rm acc} = \frac{N}{\dot N} = \frac{R_{\rm coll}}{R_{\rm sg}}\;\frac{1}{\dot N}
\label{tacc}
\end{equation}

In a quasi steady state, the surface density profile, $\Sigma(R)$, is a
strongly peaked function for these discs. We have, const$=\dot M \sim
\Sigma(R) \pi R^2/t_{\rm acc}$. As $t_{\rm acc}$ is independent of radius, we
obtain $\Sigma(R) \propto R^{-2}$. This is a much steeper dependence than that
for discs in which surface density is regulated by angular momentum transfer,
since viscous time typically is a strongly increasing function of radius. For
example, for a standard gas-dominated accretion disc, $\Sigma \propto
R^{-3/4}$ \citep{Shakura73}.

Consider gravitational stability of such discs. In our model the heating per
unit surface area is $Q^{+}\sim \Sigma \Omega^2 R^2 t_{\rm acc}^{-1}$. The
radiation flux emerging from the disc is $\sigma T^4 [\tau + 1/\tau]^{-1}$,
where $T$ is disc midplane temperature and $\tau$ is the optical depth of the
disc. Since $[\tau + 1/\tau] > 1$ for any $\tau$, we can estimate the disc
temperature to be
\begin{equation}
T > \left[\frac{\Sigma \Omega^2 R^2}{\sigma t_{\rm acc}}\right]^{1/4}\;.
\end{equation}
Further, if we take into account gas pressure only, we obtain the minimum
scale height of the disc as $H = c_s \Omega^{-1} = (kT/\mu)^{1/2}
\Omega^{-1}$. Requiring $M_{\rm disk} \approx \Sigma R^2 \simgt \mbh (H/R)$,
we arrive at the minimum disc mass required for gravitational instability in
our model:
\begin{equation}
\frac{M_{\rm sg}}{\mbh} \simgt \left[\frac{k^4 R}{G^{3} \mu^4 \mbh^2 \sigma
  t_{\rm acc}}\right]^{1/7} \approx 0.002 \; \left[\frac{R_{\rm pc}}{M_8^2
  t_5}\right]^{1/7}\;,
\label{msg}
\end{equation}
where $M_8 = \mbh/10^8 \msun$, $t_5 = t_{\rm acc}/10^5$~years. In the latter
we set $\mu = m_p$. Equation \ref{msg} allows us to estimate the minimum
accretion rate at which these discs would become self-gravitating\footnote{a
fin ate disc opacity and the radiation contribution to the disc pressure, both
neglected above, will make this only limit higher}:
\begin{equation}
\dot M_{\rm sg} = \frac{M_{\rm sg}}{t_{\rm acc}} \simgt
2\frac{\msun}{\hbox{year}} \; \left[\frac{R_{\rm pc} M_8^5}{
  t_5^8}\right]^{1/7}\;.
\end{equation}
Note that the dependence on the radius is quite weak here. Apparently, such
externally fed accretion flows are able to deliver accretion rates of the order of
the Eddington accretion rate of the SMBH if the time scale on which the
angular momentum of the incoming gas fluctuates is shorter than 
\begin{equation}
t_{\rm acc} \simlt t_{\rm sto} \sim 10^5 \; \hbox{years}\;.
\end{equation}
This time scale can be much shorter than the disc viscous time, $t_{\rm visc}
\sim \alpha^{-1} (R/H)^2 \Omega^{-1}$ at parsec distances from the SMBH
\cite{Shakura73}.

We can estimate the maximum outer radius of the externally forced
discs. Requiring that the time scale for a significant angular momentum change
$t_{\rm sto}$ to be larger than the local dynamical time given by equation
\ref{tdyn}, we see that the disc cannot be larger than about 10 pc. Another
way to put this result is to say that any disc larger than this will
necessarily be star forming.

\subsection{Inner viscous disc}\label{sec:inner}

At small radii, the viscous time is shorter than $t_{\rm acc}$. Within that
region, the inflow is viscous rather than externally driven. The inner radius
of the forced discs can thus be estimated as
\begin{equation}
R_{\rm in} \approx 0.01\;\hbox{pc}\; \alpha_{0.1}^{2/3} h_{-2}^{4/3}\; t_5^{2/3}
M_8^{1/3}\;,
\label{rin}
\end{equation}
where $h_2 = 100 H/R$ is the aspect ratio of the inner disc.
This is an estimate only, as hydrogen ionisation instability might in
principle be important for this region \citep[e.g.,][]{SE97} and modulate the
accretion rate onto the SMBH.

Lodato \& Pringle (2007) show that, in a diffusive regime, warp propagation
occurs on a time scale shorter than the viscous time by a factor
$\alpha_2/\alpha$, where $\alpha_2$ is the ``warp diffusion viscosity
coefficient''. They find that $\alpha_2$ has a maximum value of a few which is
attained for strongly warped accretion discs. Hence, in the context of our
model, the inner region of radial size
\begin{equation}
R_{\rm flat} \approx 0.1\;\hbox{pc}\; \left(\frac{\alpha_{2}}{3}\right)^{2/3}
h_{-2}^{4/3}\; t_5^{2/3} M_8^{1/3}\;,
\label{rflat}
\end{equation}
is able to flatten out due to local viscous forces. The inner flow then
consists of the innermost zone ($R < R_{\rm in}$) where the flow is flat and
viscosity mediates the inflow, and the outer zone ($R_{\rm in} < R < R_{\rm
flat}$), where the disc is also flat but the inflow is driven by the same
mechanism as in the larger forced disc. The orientation of the inner disc $R <
R_{\rm flat}$ coincides with the that of the innermost part of the larger
forced disc. 

Note that $R_{\rm flat}$ sets a velocity scale -- the Kepler velocity at that
radius -- of $v \simeq$~few~$\times 10^3$ km s$^{-1}$. Since the inner disc is
flat, it presents a very small solid angle to the AGN illumination. On the
other hand, the outer disc might be much better exposed to the AGN. It is then
possible that the transition between the inner flat and the outer warped disc
will be a site for broad AGN optical and UV lines. We would then predict, in
the context of our model, that the broad line features should have velocity
widths of order of a few thousand km s$^{-1}$.

\section{Implications}

\paragraph*{Avoiding the self-gravity catastrophe.} The picture of accretion
proposed above suggests that stochastically fed AGN accretion discs can extend
beyond the self-gravity radius $R_{\rm sg}$. These discs are not subject to
the self-gravity catastrophe as long as the gas feeding these has a random
sense of angular momentum fluctuating on time scales $t \le t_{\rm sto}$. This
then ensures that super-massive black holes can grow via gas accretion, as
required by the observations \citep{Yu02}. 

Note that our model is related to the proposals made by \cite{Goodman03} and
especially \cite{KingPringle07} that material feeding AGN comes in shots
directly impacting the self-gravity radius $R_{\rm sg}$. The latter is very
small, $R_{\rm sg} \sim 0.02$ pc. We therefore think that the picture proposed
here is more likely. The incoming clouds do not have to carry such a small
angular momentum, and they first form a disc that can be much larger than
$R_{\rm sg}$.

\paragraph*{Nuclear star formation.} In our model, star formation in the direct vicinity 
of a SMBH is a function of not only the gaseous mass deposited there but also
the manner in which that mass arrived there. In particular, contrary to
steady-state viscous accretion flows, it is feasible to have mainly accretion
at a high mass deposition rate, but mainly star formation at lower rates. An
example of the latter situation could be the central parsec of our Galaxy,
where two young stellar rings have apparently formed in situ about $\sim 6$
Million years ago \citep{Levin03,Genzel03a}. A reasonable explanation of this
is an infall of two clouds with mass of a few thousand to $\sim 10^4 \msun$
\citep{NC05,Paumard06}. Coeval infall of several more clouds of this type
from random directions could have led to more angular momentum cancellation
and mainly accretion of gas on \sgra\ instead of star formation. 

There is no one-to-one relation between AGN activity and nuclear starbursts in
this picture, as it is possible {\em in principle} to have either one
separately. On the other hand, AGN feeding in our model depends on
availability of gas clouds on low angular momentum orbits. Star formation
feedback (outflows and supernovae) may be encouraging such orbits via adding
large random velocity kicks to gas clouds in vicinity. Thus, statistically it
is more likely that starbursts and AGN would be linked to one another.

\paragraph*{Masing and/or star forming rings in nearby AGN.}

If SMBHs are fed by grand design discs, the surface density distribution is
normally a continuous well behaved function, such as a power-law. In contrast
to that, if accretion is stochastic, then the disc surface density
distribution does not have to be a smooth function, particularly at lower
cloud deposition rates. In particular, rings with radial extent $\Delta R
\simlt R$ might result. Observationally this might be relevant to masing discs
in nearby AGN, where the radial extent of the emitting region is usually
narrow. If rings are massive enough, stellar discs with well defined inner and
outer edges may form in this way.

\paragraph*{Random SMBH jet orientation, faster early SMBH growth.} 
As pointed out by \cite{KingPringle07}, the spin of a black hole fed by
deposition of clouds with randomly directed angular momentum will be
frequently misaligned with that of the inner accretion disc. This leads to a
smaller radiative efficiency and a faster black hole growth, helping to
explain the heavy-weight champion SMBHs observed already at high redshifts
(tbd). Furthermore, \cite{SchmittEtal02} pointed out that jets in radio galaxies
seem to be oriented (almost) randomly with respect to dust discs of these
galaxies. This would be natural in our model.

\paragraph*{The $\mbh$-$M_{\rm bulge}$ correlation.} Models explaining the 
correlation between the observed SMBH masses and the bulge masses hosting them
as arising due to SMBH accretion feedback seem to be promising
\citep{King03}. These models postulate that gas located throughout the bulge
can be overheated or swept away by the SMBH feedback. If SMBH were fed from a
large scale massive gaseous disc with a small $H/R$ ratio, it would be very
hard to affect that gas reservoir.

The difficulty in expelling a flat disc is two fold: (i) Firstly, the column
depth of the gaseous disc with mass $M_d$ through the midplane of the disc is
$\sim M_{d}/(RH)$. This is $R/H$ times higher than the column depth of
spherically distributed material with same mass, $\sim M/R^2$. Hence it is
harder to affect the disc with feedback of any type. (ii) Secondly, the SMBH
feedback may well be collimated and oriented perpendicular to the inner
accretion disc. If the inner disc is oriented same way as the much larger disc
in the bulge, then the feedback may miss the disc altogether. A numerical
illustration of these principles is provided by the recent calculation of the
radiation field during build up of a high mass protostar by
\cite{Krumholz05}. These simulations show that if outflow is collimated,
then the radiation field is also collimated and therefore affects the infalling
material significantly less than previously thought. 

Thus, there does not seems to be a way to expel either the disc from the bulge
or shut off SMBH feeding if this originates in a massive flat disc.  On the
other hand, if material feeding the SMBH is distributed quasi-isotropically
throughout the bulge, and if the direction of the SMBH spin (and feedback)
randomly fluctuates, then the arguments made by \cite{King03} apply and may
explain the $\mbh$-$M_{\rm bulge}$ correlation.

\paragraph*{Absence of a correlation between AGN activity and presence of bars.} 
If SMBH is not fed by the large scale grand design gaseous features such as
spirals or bars, then the latter have no bearing on AGN activity, as observed.

\paragraph*{Importance of mergers in AGN feeding.} Here we have argued that SMBH
fuelling is driven by infall of gas on nearly radial trajectories. For the gas
at $R\sim R_{\rm bulge}$ to assume such an orbit, the angular momentum of
(only some!) clouds needs to cancel out almost completely by cloud-cloud
collisions, for example. A way of putting much more gas on such orbits would
be to have a major merger, in which the gaseous discs of the galaxies collide
bodily. Such collisions probably generate SMBH feeding rates well in excess of
the Eddington rate since the gaseous discs can be orders of magnitude more
massive than the SMBHs.

\paragraph*{Nuclear star clusters and SMBH in dwarf galaxies.} Galaxies going 
through fewer mergers channel less gas on nearly radial orbits. Central black
holes in these galaxies are thus relatively more ``fuel starved'' than their
cousins in galaxies experiencing more mergers. SMBHs in such ``no mergers''
galaxies could then be underweight compared to their expected $\mbh$-$M_{\rm
bulge}$ or $\mbh$-$\sigma$ mass. If smaller galaxies go through fewer mergers,
as current cosmological simulations imply, then it is these galaxies, i.e.,
dwarf spheroidals, that are most strongly affected by this argument.  The fuel
stalled in the central region of the galaxy because of insufficient angular
momentum cancellation can be used up in nuclear star formation. These ideas
might be relevant to the recently claimed dichotomy of nuclear star clusters
and SMBHs in observations \citep[e.g.,][]{Wehner06}, with the former objects
present mainly in low mass galaxies, and with dwarf spheroidals possibly
lacking SMBHs.

\paragraph*{The final parsec problem for SMBH mergers.} It is found that
central black holes in a major merger always find their way into the central
part of the resulting galaxy by dynamical friction
\citep[e.g.,][]{BegelmanEtal80}. These black holes then form a binary. The SMBH
binary continues to shrink by expelling stars. However, when the binary
separation approaches a few parsec, interactions with the stellar background
become too inefficient.  If gas continues to pile up in the inner few parsecs
of the galaxy through stochastic cloud deposition, the SMBH pair can do work
on this supposedly unlimited gas supply instead of stars, and continue to
shrink until gravitational radiation takes over \citep[for related ideas
see][]{EscalaEtal05}.

\paragraph*{Unification and obscuration schemes of AGN.} Warped accretion discs
have been claimed to be important in the obscuration of the inner regions of
type II AGNs \citep{Nayakshin05}. Stochastically fed accretion discs are
generically strongly warped because different mass ``shots'' will likely have
not only random angular momentum orientation and also different
circularisation radii. Further, it can be shown that time scales for
flattening the warps can be much longer than $t_{\rm sto}$ (except for the
inner disc, see \S \ref{sec:inner}).

We have shown that our disc model is gravitationally stable for disc masses
less than $M_{\rm sg} \sim 0.002 M_{\rm bh}$ (see equation \ref{msg}).  The
column depth of a such a disc is then
\begin{equation}
\Sigma \sim \frac{M_{\rm disc}}{\pi R^2} \sim 20\; M_8^{5/7}\ R_{\rm pc}^{-13/7}
t_5^{-1/7}\;,
\label{sigma}
\end{equation}
Thus, if these discs are strongly warped, they could form Compton-thick
absorbers with size of at most a few parsec. This agrees with estimates for
the obscuring medium inferred in AGN.

\paragraph*{Absence of type II LLAGN.} In our picture, Low Luminosity AGN 
would deposit mass at a low rate or always with the same angular momentum. In
either case this would imply a disc passively circling the SMBH at the
circularisation radius or perhaps forming stars if the disc becomes too
massive. In the first case, the disc may well have enough time to flatten out
by viscous or gravitational torques, whereas in the other the disc might be
consumed by star formation too rapidly to provide a large enough obscuring
column depth. therefore, in general we predict that LLAGN must be much less
obscured than brighter AGN such as Seyfert Galaxies.

\section{Discussion and Conclusions}

In this paper we have considered SMBH feeding
in galaxies. We suggest that due the extremely long time scales for
angular momentum transfer and the loss of gas to star formation mean that large
scale gas discs do not contribute directly to SMBH growth. Instead, we argue
that SMBHs are fuelled by low angular momentum gas. Our model is
closely related to suggestions made by \cite{Goodman03} and
\cite{KingPringle07}, as the core idea is that the angular momentum of the
incoming gas is small in a time-averaged sense. However, our model has
additional observational implications. For example, warped ``externally
forced'' accretion discs can extend to scales of up to 10 pc (see the end of
\S \ref{sec:forced}), which is much larger than the self-gravity radius
$R_{\rm sg} \sim 0.01-0.1$~pc.

One question that we do not address here is whether real galaxies feed enough
gas on nearly radial trajectories to support the growth of their SMBHs. This
requires numerical simulations of a significant dynamic range and physical
complexity.

\bibliographystyle{mnras} 

\end{document}